\documentclass[traditabstract,longauth,letter]{aa}  
\usepackage{graphicx}
\usepackage{txfonts}
\usepackage{natbib}
\newcommand{\Lsun}{{\hbox {L$_\odot$}}}
\newcommand{\Msun}{{\hbox {M$_\odot$}}}

\newcommand{\refmod}{{\hbox {$M_{\mathrm{ref}}$}}}
\def\t#1#2#3#4#5#6{{\hbox {$#1_{#2#3}\!\rightarrow\!#4_{#5#6}$}}}

\def\13co{$^{13}$CO}
\def\c18o{C$^{18}$O}

\begin{document}
   \title{Herschel\thanks{Herschel is an ESA space observatory with science
       instruments provided by European-led Principal Investigator consortia
       and with important participation from NASA.} observations of water
     vapour in Markarian 231}

   \author{E. Gonz\'alez-Alfonso \inst{1}, 
          J. Fischer \inst{2},
          K. Isaak \inst{3},
          A. Rykala \inst{3},
          G. Savini \inst{4},
          M. Spaans \inst{5},
          P. van der Werf \inst{6},
          R. Meijerink \inst{6},
          F. P. Israel \inst{6},
          A. F. Loenen \inst{6},
          C. Vlahakis \inst{6}, 
          H. A. Smith\inst{7}, 
          V. Charmandaris \inst{8,21},
          S. Aalto \inst{9},
          C. Henkel \inst{10},
          A. Wei{\ss} \inst{10},
          F. Walter \inst{11},          
          T. R. Greve \inst{11,12},
          J. Mart\'{\i}n-Pintado \inst{13},
          D. A. Naylor \inst{14},
          L. Spinoglio \inst{15},
          S. Veilleux \inst{16},
          A. I. Harris \inst{16},
          L. Armus \inst{17},
          S. Lord \inst{17},
          J. Mazzarella \inst{17},
          E. M. Xilouris \inst{18},
          D. B. Sanders \inst{19}, 
          K. M. Dasyra \inst{20},
          M. C. Wiedner \inst{21}, 
          C. Kramer \inst{22},
          P. P. Papadopoulos \inst{23},
          G. J. Stacey \inst{24},
          A. S. Evans\inst{25}, 
          \and
          Y. Gao \inst{26}
          }

   \institute{Universidad de Alcal\'a de Henares, Departamento de F\'{\i}sica, 
Campus Universitario, E-28871 Alcal\'a de Henares, Madrid, Spain 
         \and
            Naval Research Laboratory, Remote Sensing Division, 
Washington, DC 20375, USA 
         \and
ESA Astrophysics Missions Div/ Research and Scientific Support Dept
ESTEC/SRE-SA Keplerlaan 1, NL-2201 AZ Noordwijk, The Netherlands
         \and
Department of Physics \& Astronomy, University College London,
Gower Street, London WC1E 6BT ,UK
         \and
Kapteyn Astronomical Institute, University of Groningen, P.O.
Box 800, 9700 AV, Groningen, the Netherlands
         \and
Leiden Observatory, Leiden University, P.O. Box 9513, 2300 RA Leiden, The
Netherlands 
         \and
Harvard-Smithsonian Center for Astrophysics,
    60 Garden Street, Cambridge, MA 02138, USA
         \and
University of Crete, Department of Physics, GR-71003, Heraklion, Greece
         \and
Onsala Space Observatory, Chalmers University of Technology, 439 92 Onsala,
Sweden 
         \and
MPIfR, Auf dem H\"ugel 69, 53121 Bonn, Germany
         \and
Max-Planck-Institut f\"ur Astronomie, K\"onigstuhl 17, D-69117 Heidelberg,
Germany
         \and
Dark Cosmology Centre, Niels Bohr Institute, University of Copenhagen,
Juliane Maries Vej 30, 2100 Copenhagen, Denmark
         \and
Centro de Astrobiolog\'{\i}a (CSIC-INTA), Ctra de Torrej\'on a Ajalvir, km 4,
28850 Torrej\'on de Ardoz, Madrid, Spain
         \and
Space Astronomy Division, Institute for Space Imaging Science,
Department of Physics and Astronomy, University of Lethbridge,
Lethbridge, Alberta Canada, T1K 3M4
         \and
Istituto di Fisica dello Spazio Interplanetario, CNR
via Fosso del Cavaliere 100, I-00133 Roma, Italy
         \and
Department of Astronomy, University of Maryland, College
Park, MD 20742 USA
         \and
IPAC, California Institute of Technology, MS 100-22,
Pasadena, CA 91125, USA
         \and
Institute of Astronomy and Astrophysics, National Observatory
of Athens, P. Penteli, GR-15236 Athens, Greece
         \and
University of Hawaii, Institute for Astronomy, 2680 Woodlawn Drive,
Honolulu, HI  96822,  USA
         \and
Laboratoire AIM, CEA/DSM - CNRS - Universit\'e Paris Diderot,
Irfu/Service d'Astrophysique, CEA Saclay, Orme des Merisiers, 91191
Gif sur Yvette Cedex, France
         \and
LERMA, Observatoire de Paris, CNRS, 61 Av. de l'Observatoire,
75014 Paris, France
         \and
Instituto Radioastronomia Milimetrica (IRAM),
Av. Divina Pastora 7, Nucleo Central, E-18012 Granada, Spain
         \and
Argelander Institut fuer Astronomy,
Auf dem Huegel 71, D-53121, Germany
         \and
Department of Astronomy, Cornell University, Ithaca, NY, USA
         \and
Department of Astronomy, University of Virginia, 530 
McCormick Road, Charlottesville, VA 22904, USA
         \and
Purple Mountain Observatory, Chinese Academy of Sciences, 2 West Beijing Road,
Nanjing 210008, China
}


 
   \authorrunning{Gonz\'alez-Alfonso et al.}
   \titlerunning{Herschel observations of H$_2$O in Mrk 231}

  \abstract
   {The Ultra Luminous InfraRed Galaxy (ULIRG) Mrk\,231 reveals up to seven
     rotational lines of water (H$_2$O)
     in emission, including a very high-lying ($E_{\mathrm{upper}}=640$
     K) line detected at a 4$\sigma$ level, within the Herschel/SPIRE 
     wavelength range ($190 < \lambda (\mathrm{\mu m})< 640$), whereas PACS
     observations show one H$_2$O line at 78 $\mu$m in absorption, as found
     for other H$_2$O lines previously detected by ISO.
   The absorption/emission dichotomy is caused by the
     pumping of the rotational levels by far-infrared radiation emitted by
     dust, and subsequent relaxation through lines at longer wavelengths,
     which allows us to estimate both the column
     density of H$_2$O and the general characteristics of the underlying
     far-infrared continuum source.  
   Radiative transfer models including excitation through both absorption of
     far-infrared radiation emitted by dust and collisions are used to
     calculate the equilibrium level populations of H$_2$O and the
     corresponding line fluxes.
     The highest-lying H$_2$O lines detected in emission, with levels at
     $300-640$ K above the ground state, indicate that the source of
     far-infrared radiation responsible for the pumping is compact
     (radius$=110-180$ pc) and warm ($T_{\mathrm{dust}}=85-95$ K),
     accounting for at least $45$\% of the bolometric luminosity. The high
     column density, $N(\mathrm{H_2O})\sim5\times10^{17}$ cm$^{-2}$, found
     in this nuclear component, is most probably the consequence of
     shocks/cosmic rays, an XDR chemistry, and/or an ``undepleted chemistry''
     where grain mantles are evaporated. A more extended region, presumably
     the inner region of the 1-kpc disk observed in other molecular species,
     could contribute to the flux observed in low-lying H$_2$O lines through
     dense hot cores, and/or shocks. The H$_2$O 78 $\mu$m line observed with
     PACS shows hints of a blue-shifted wing seen in absorption, possibly
     indicating the occurrence of H$_2$O in the prominent outflow detected  in
     OH (Fischer et al., this volume). Additional PACS/HIFI observations of
     H$_2$O lines are required to constrain the kinematics of the nuclear
     component, as well as the distribution of H$_2$O relative to the 
     warm dust.}


   \maketitle
%

\section{Introduction}

   One key question in the study of composite infrared (IR) merging galaxies
   and quasi-stellar objects (QSOs) is what fraction of their luminosity is 
   generated in the nuclear region ($<200$ pc) associated with the Active
   Galactic Nucleus (AGN) and a possible extreme nuclear starburst, and what
   fraction arises from a more extended kpc-scale starburst
   \citep[e.g.][]{arm07,vei09}. The 
   ULIRG Markarian 231 (Mrk 231) is the most luminous ($L\sim4\times10^{12}$
   \Lsun) galaxy in the local Universe ($z<0.1$), 
   and thus provides a unique template for such studies.
   Since the bulk of
   the luminosity in ULIRGs arises at far-IR wavelengths, where
   sub-arc-second resolution observations are not available, an alternative
   technique is required to constrain the compactness of the far-IR
   emission and its physical origin.
 
   In a previous work based on observations with the \emph{Infrared Space
     Observatory (ISO)}, \citet[][hereafter G-A08]{gon08} have
   argued  that the observation of molecular species such as OH and H$_2$O at
   far-IR wavelengths is ideal for such a purpose, because their high-lying
   rotational levels are pumped through absorption of far-IR radiation and
   the observable excitation is then sensitive to the far-IR radiation density
   that in turn depends on the compactness of the far-IR continuum source. In
   addition, these molecular observations shed light on the dominant chemistry
   in those nuclear regions.  
   G-A08 reported the ISO detection of 3 high-lying H$_2$O lines, relevant
   upper limits over the entire ISO  
   spectrum, and also high-lying OH lines, indicating the 
   ocurrence of a compact-luminous far-IR component.

   With their high sensitivity, spectral resolution, and wavelength
   coverage, the \emph{Herschel} \citep{pil10} instruments are ideal for
   extending our previous 
   study to additional key lines in the far-IR/submillimeter.
   As part of the HerCULES Key Programme \citep[see][this volume, hereafter
   vdW10]{vdw10}, we report in this Letter the
   Herschel SPIRE/PACS \citep{gri10,pog10} detection and first analysis of
   several H$_2$O lines in 
   Mrk 231, which supports the conclusions of G-A08 and gives
   additional clues to the origin of H$_2$O in this ULIRG. We adopt a
   distance to Mrk 231 of 192 Mpc ($H_0=70$ km s$^{-1}$ Mpc$^{-1}$,
   $\Omega_{\Lambda}=0.73$, and $z=0.04217$).


\section{Observations}

   \begin{figure}
   \centering
   \includegraphics[width=8.0cm]{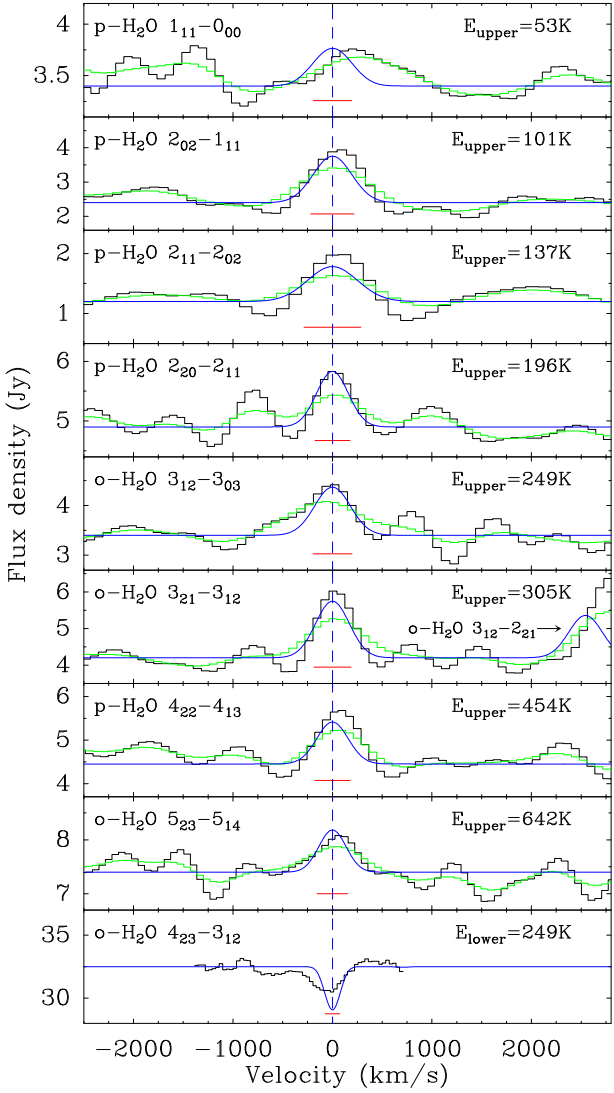}
   \caption{Comparison between the observed spectra (black/green histograms:
     unapodized/apodized spectrum, see vdW10) and results 
    for the reference model discussed in \S3\ (dark blue lines; a
    Gaussian instrumental line shape with 
    $\mathrm{FWHM}=0.048$ cm$^{-1}$ is used for simplicity). The \t312221\
    line, shown in the same panel as the \t321312\ line, is blended with
    $^{12}$CO (10-9) (vdW10). The red segment 
    in each panel indicates the FWHM of an unresolved line. The velocity scale
    has been calculated with respect to the systemic redshift of $z=0.04217$.}
              \label{Figspectra}
    \end{figure}

   \begin{figure}[ht]
   \centering
   \includegraphics[width=7.5cm]{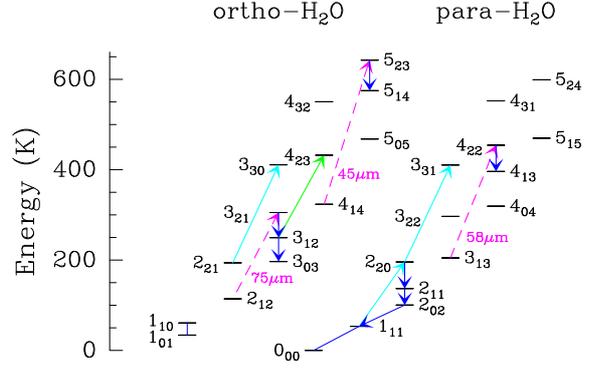}
   \caption{Energy level diagram for H$_2$O, showing the detected/undetected
     (blue arrows/lines) lines with SPIRE, the line detected with PACS
     (green) and those detected by ISO (light blue).
     Dashed red arrows indicate the main pumping paths for the high-lying
     lines observed with SPIRE. Upward (downward) arrows: absorption
     (emission) lines.}
              \label{Figlevels}
    \end{figure}

The SPIRE FTS observations of Mrk 231 (vdW10) were conducted on December 9th,
2009. The PACS observation of one H$_2$O line was conducted on 
November 8th, 2009, as part of the SHINING Key Programme, and kindly provided
for the present study. Details on data reduction, calibration, and line
extraction are given in vdW10 and \citet{swi10}. Excerpts of the spectrum
around the H$_2$O lines are displayed in 
Fig.~\ref{Figspectra}, and an energy level diagram indicating the lines
detected with SPIRE, the one detected with PACS, and those detected with ISO
is shown in Fig.~\ref{Figlevels}. Table~\ref{tab:fluxes} lists
the line fluxes. Figure~\ref{Figspectra} also shows the results of our
reference model, discussed below.

H$_2^{18}$O could be marginally detected at 250.0 $\mu$m (\t220211) (see
vdW10); however, this identification should be 
confirmed as the feature is shifted by 150 km s$^{-1}$ from the nominal
line wavelength. Confirmation of H$_2^{18}$O would be important as its
presence would support the strong enhancement of $^{18}$O in Mrk 231
derived from $^{18}$OH observations \citep[][this issue, hereafter
F10]{fis10}. A number of the H$_2$O lines in Fig.~\ref{Figspectra} 
are blended with the C$^{18}$O lines $9-8$ (303.57 $\mu$m), $10-9$ (273.24
$\mu$m), and $11-10$ (248.43 $\mu$m). 
Contamination by C$^{18}$O is minimal, however, 
as the lower-lying $6-5$, $7-6$, and $8-7$ lines are not
detected. PACS observations show a broad absorption feature at 121.7
$\mu$m, nearly coincident with the H$_2$O \t432423\ line (F10). However, 
this feature is probably contaminated by HF $(2-1)$ at the same
wavelength, as the $1-0$ line is detected with SPIRE (vdW10); 
HF has been previously detected towards the Galactic star forming region Sgr
B2 \citep{neu97}.  
Therefore the 121.7 $\mu$m feature is not used for the H$_2$O analysis below.

Of particular interest is the detection of the very high-lying \t523514\
H$_2$O line at a 4$\sigma$ level, which we have verified by reprocessing
the data with no correction applied for the instrument spectral efficiency and
by comparing these data with reduced observations of the dark sky
over the same spectral range. 
The ground state lines (p-H$_2$O \t111000\ in Fig.~\ref{Figspectra} and
o-H$_2$O \t110101) are not detected. 
The lines detected with SPIRE are all
in emission and peak at central velocities, in contrast to
the low-lying OH lines 
that show P-Cygni profiles characteristic of an extreme molecular outflow
(F10). The red horizontal lines indicate the
FWHM of an unresolved line, and show that the H$_2$O lines detected with SPIRE
are barely resolved. The o-H$_2$O \t423312\ line detected with the
higher spectral resolution of PACS is in absorption and well
resolved, showing a central main body and, apparently, a
relatively weak blue-shifted wing extending up to $-800$ km s$^{-1}$ and 
possible low level emission from the receding gas. These wings should be
confirmed with additional observations of 
H$_2$O absorption lines, as the limited wavelength coverage of the \t423312\
line makes the adopted baseline uncertain. The detection would indicate that
H$_2$O also participates in the prominent outflow detected in OH (F10).

While the shape of the \t423312\ absorption line shows the
centroid of the main body slightly blue-shifted (by $\sim-70$ km s$^{-1}$), 
some lines observed in emission tend to show, on the contrary, a slight
red-shift of their centroid (up to $100$ km s$^{-1}$). This effect could be
related to systematic motions, i.e. a low-velocity nuclear-scale outflow, and
will be explored in the future with additional high spectral resolution
observations.

   \begin{table}
      \caption[]{Observed and modeled line fluxes.}
         \label{tab:fluxes}
          \begin{tabular}{lccc}   
            \hline
            \noalign{\smallskip}
            Line  &  $\lambda$ & $\mathrm{Flux}^{\mathrm{a}}$ & Model \\ 
                & ($\mu$m)  & ($\mathrm{Jy\,\, km\,\,s^{-1}}$) & 
                 ($\mathrm{Jy\,\,km\,\,s^{-1}}$) \\
            \noalign{\smallskip}
            \hline
            \noalign{\smallskip}
            p-H$_2$O \t111000 &  $269.27$  & $< 300$   &   $177$ \\
            o-H$_2$O \t110101 &  $538.29$  & $< 400$   &   $146$ \\
            p-H$_2$O \t202111 &  $303.46$  & $718$ ($110$) &  $660$ \\
            p-H$_2$O \t211202 &  $398.64$  & $415$ ($43$)  &  $365$ \\
            p-H$_2$O \t220211 &  $243.97$  & $342$ ($92$) &  $372$ \\
            o-H$_2$O \t312303 &  $273.19$  & $400$ ($130$)  & $438$ \\
            o-H$_2$O \t321312 &  $257.79$  & $631$ ($47$)  &  $657$ \\
            p-H$_2$O \t422413 &  $248.25$  & $361$ ($38$)  &  $381$ \\
            o-H$_2$O \t523514 &  $212.53$  & $287$ ($83$)  &  $269$ \\
            o-H$_2$O \t423312 &$78.74$& $-910^{\mathrm{b}}$ ($60$)  & $-664$ \\
            \noalign{\smallskip}
            \hline
         \end{tabular} 
\begin{list}{}{}
\item[$^{\mathrm{a}}$] Numbers in parenthesis indicate the estimated
  uncertainties. 
\item[$^{\mathrm{b}}$] Includes the absorption in the high-velocity
  blue-shifted wing, which accounts for $\sim-220$ $\mathrm{Jy\,\,
    km\,\,s^{-1}}$. 
\end{list}
   \end{table}

\section{Analysis}

The observed pattern of line emission cannot be explained in terms of
{\em pure} collisional excitation (G-A08). 
Adopting the collisional rates of \citet{fau07} with $T_k>200$ K\footnote{
    An LTE ortho-to-para H$_2$ ratio according to $T_k$ is assumed.} and
$n(\mathrm{H_2})=10^6-10^7$ cm$^{-3}$, and ignoring
radiative pumping, the models
that account within a factor of 2 for the high-lying \t523514\ and \t422413\
lines predict fluxes for the low-lying lines that exceed the observed values
by factors of $\gtrsim10$. Thus the observed line ratios indicate an
excitation mechanism that favours the emission in the high-lying ($\gtrsim300$
K) lines at the expense of the low-lying lines. This does {\em not} imply that 
the H$_2$O lines are {\em not} formed in warm/dense regions, 
but just that the dominant excitation mechanism for the {\em high-lying} lines
is not collisional.

\begin{figure}
   \centering
   \includegraphics[width=8.0cm]{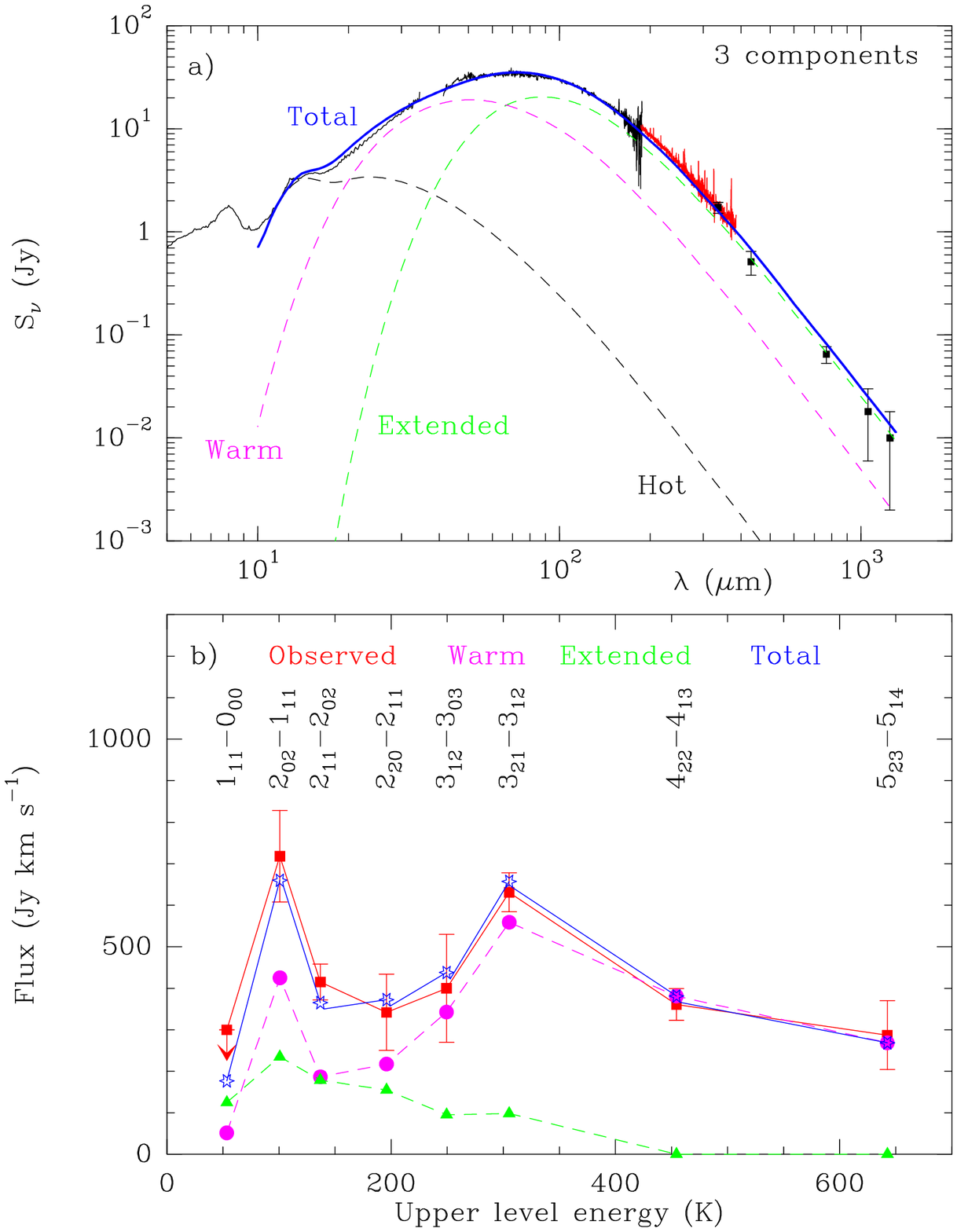}
   \caption{Comparison between observations and results for the reference
     model. a) Continuum emission from Mrk 231. Spitzer IRS
     data \citep{arm07}, ISO data (G-A08), and SPIRE data (red
     spectrum) are shown. Flux densities at 800
and 1100 $\mu$m are taken from \citet[][corrected for
non-thermal emission]{roc93}, at 450 $\mu$m from \citet{rig96}, and at 350
$\mu$m from \citet{yan07}. b) H$_2$O emission. The high-lying
     lines are generated in the warm/compact region (violet), whereas the
     extended kpc-scale starburst (green) is expected to contribute to the
     low-lying lines.
} 
              \label{Figmref}
    \end{figure}

Such a mechanism is the pumping 
through absorption of dust-emitted far-IR photons, which efficiently pumps the
high-lying \t321312$/$\t422413$/$\t523514\ lines through 
absorptions at $75.4$/$57.6$/$45.1$ $\mu$m in the strong
\t321212$/$\t422313$/$\t523414\ lines (Fig.~\ref{Figlevels}), 
of which the lower backbone levels are
preferently populated. This requires a strong continuum component 
at $30-70$ $\mu$m. However, this component cannot dominate the
emission at $>130$ $\mu$m, as it would produce strong H$_2$O absorptions in
that wavelength range that are not observed (G-A08). The data then support
the occurrence of both a warm/compact component with moderate opacity, and a
second colder component naturally associated with the more extended 1-kpc
starburst that dominates the emission at long wavelengths. Our proposed SED
decomposition is shown in Fig.~\ref{Figmref}a, and defines the {\em reference
  model} (\refmod) with parameters as listed in Table~\ref{tab:model}: 
$(i)$ a hot component ($H_C$) with $T_{\mathrm{dust}}=150-400$ K 
dominates the emission at $\lambda<20$ $\mu$m; $(ii)$ 
a warm ($95$ K) and compact ($R=120$ pc) component ($W_C$)
dominates at $20\,\mu\mathrm{m}<\lambda<70\,\mu\mathrm{m}$; $(iii)$  
an extended 1-kpc component ($E_C$), with $T_{\mathrm{dust}}=40$ K, accounts
for most of the continuum at $\lambda>70$ $\mu$m.  
The $W_C$, with $L_W\sim1.9\times10^{12}$ \Lsun, is responsible
for the observed high-lying H$_2$O line emission.

Calculations for H$_2$O were carried out using the code described in
\cite{gon99}. In \refmod\ (Table~\ref{tab:model}),
line broadening is caused by microturbulence. We
have adopted a ``mixed'' approach (i.e. the H$_2$O
molecules are evenly mixed with dust, G-A08), discussed
below. An ortho-to-para H$_2$O abundance ratio of 3 is assumed.  
For those H$_2$O lines observed in emission, Fig.~\ref{Figmref}b compares
the expected fluxes from the $W_C$ (in violet) with the observed
fluxes (in red). Collisional excitation is included with gas at
$T_k=150$ K and $n(\mathrm{H_2})=1.5\times10^6$ cm$^{-3}$ but,
even for these shock-like conditions\footnote{The average
  densities are much lower than those adopted for collisional excitation, 
$<\!\!n(\mathrm{H_2})\!\!>\approx200$ and 2000 cm$^{-3}$ for $E_C$ and $W_C$.}, 
it has a low 
effect on the H$_2$O level populations and line fluxes as these are mostly 
determined by the strong radiation field. The high-lying lines are reproduced
with the $W_C$, but there is a model deficit of emission from the
low-lying lines. This deficit indicates that the $E_C$ contributes to those
low-lying lines (green); both radiative and collisional excitation, the latter
significant for $E_{\mathrm{upper}}\leq200$ K, are included in the model.

The high-lying \t422413\ and \t523514\ lines are (nearly)
optically thin in the $W_C$, so that their expected fluxes are sensitive to
$N(\mathrm{H_2O})$. Table~\ref{tab:model} shows that, despite
the strong radiation field in this region, a high $N(\mathrm{H_2O})$ is
required to account for the observed fluxes. Assuming a gas-to-dust mass
ratio of 100, the average H$_2$O abundance relative to H$_2$ in this $W_C$ is
$0.8\times10^{-6}$.

The H$_2$O line at 78.7 $\mu$m observed by PACS is consistently predicted in
absorption, and its flux is reasonably reproduced (Table~\ref{tab:fluxes}),
given that the blue-shifted wing is not modeled. However, the observed line
shape suggests systematic motions that are not included in our model. In
general, we expect that lines with $\lambda\lesssim120$ $\mu$m are observed in
absorption, while lines with $\lambda\gtrsim120$ $\mu$m are in emission 
(with exceptions due to the pumping details and level energies).

Concerning the ISO lines (G-A08),
\refmod\ also matches the absorption in the \t220111\ line at $101$
$\mu$m, but underestimates the absorption in the \t330221\ and \t331220\ lines
at $66.4$ and $67.1$ $\mu$m by a factor of 2. As mentioned above, our model
uses a mixed approach, which implies that for given values of
$N(\mathrm{H_2O})$, the absorbing H$_2$O lines are relatively weak, as
molecules located deep inside the source do not contribute to the
absorption features. Conversely, if a screen approach is adopted (i.e. a
H$_2$O shell surrounds the continuum source), the absorption lines become much
stronger. In both approaches, 
the $N(\mathrm{H_2O})$ required to match the lines observed with
SPIRE are similar. Thus an analysis that combines absorption and
emission lines is a powerful tool to establish the distribution of H$_2$O
relative to the warm dust responsible for the excitation. The screen
version of \refmod\ yields absorption in \t330221, \t331220, and other
transitions that overestimate the observed values or upper limits. Therefore,
our preliminary result is that a combination of both the mixed and screen
scenarios best describes that observed data, with the mixed version
favoured. Nevertheless, it remains unclear what fraction of the observed
\t330221\ and \t331220\ absorption arises from the outflow detected in OH
(F10).

By increasing $R_C$, the radius of $W_C$, to 170 pc, and
keeping $L_W$ constant, the \t321312/\t423312\ line strengths become
overestimated by 
$50/25$\%, and the \t211202/\t523514\ intensities are underestimated by
30\%;  given the simplicity of our spherically symmetric models, 
we estimate a size for $W_C$ in the range $R_C=110-180$ pc. A lower 
limit for the luminosity arising from $W_C$ is estimated by decreasing
$T_{\mathrm{dust}}$ to $85-80$ K and increasing $N(\mathrm{H_2O})$ to
$\gtrsim10^{18}$ cm$^{-2}$, which results in too weak \t423312\ and ISO
absorption lines. We estimate that the  
mid- and far-IR emissions from the nuclear region account for more than 45\% 
of the bolometric luminosity; observations at $60-200$ $\mu$m are required
to better constrain that value and to establish a firm upper limit.
Results are more uncertain for $E_C$, and we may expect that its contribution
to the H$_2$O emission arises from its innermost region.

   \begin{table}
      \caption[]{Parameters of the reference model (\refmod).}
         \label{tab:model}
          \begin{tabular}{lccc}   
            \hline
            \noalign{\smallskip}
     Component$\rightarrow$  & Hot$^{\mathrm{a}}$ ($H_C$) & Warm ($W_C$) &
     Extended ($E_C$)\\  
            \noalign{\smallskip}
            \hline
            \noalign{\smallskip}
        Radius (pc)            &   $23$       &  $120$   &    $610$ \\
        $T_{\mathrm{dust}}$ (K)  &  $400-150$  & $95$   &   $41$ \\
      $\tau_{\mathrm{100\,\mu m}}$ &  $0.4$     & $1.0$   &   $0.5$ \\
 $L$ (\Lsun) & $7.5\times10^{11}$ & $1.9\times10^{12}$ & $9.6\times10^{11}$ \\
Gas Mass$^{\mathrm{b}}$ (\Msun) & $1.9\times10^{6}$ & $5.9\times10^{8}$ &
$7.7\times10^{9}$ \\ 
$N(\mathrm{H_2O})$ (cm$^{-2}$) & $-$ & $5.2\times10^{17}$ & $2.0\times10^{16}$ \\
$V_{\mathrm{turb}}$ (km s$^{-1}$) & $-$ & $60^{\mathrm{d}}$ & $40$ \\
$n(\mathrm{H_2})^{\mathrm{c}}$ (cm$^{-3}$) & $-$ & $1.5\times10^6$ &
 $5\times10^5$ \\ 
$T_{\mathrm{gas}}^{\mathrm{c}}$ (K)   & $-$ & $150$ & $100$ \\ 
            \noalign{\smallskip}
            \hline
         \end{tabular} 
\begin{list}{}{}
\item[$^{\mathrm{a}}$] The hot component does not contribute to the H$_2$O
  emission.
\item[$^{\mathrm{b}}$] A gas-to-dust mass ratio of 100 is assumed.
\item[$^{\mathrm{c}}$] These parameters are not well determined for $W_C$,
  as the excitation is dominated by radiative pumping.
\item[$^{\mathrm{d}}$] From \cite{dow98}.
\end{list}
   \end{table}

\section{Discussion}

The extreme nature of the nuclear region in Mrk 231 is well illustrated by
comparing its SPIRE spectrum with that of the Orion Bar 
\citep[][this issue]{hab10}, the prototypical Galactic PDR. The Orion Bar
spectrum shows CO lines a factor of $\gtrsim50$ stronger than the H$_2$O lines,
while in Mrk 231 the H$_2$O and CO lines have comparable strengths
(vdW10). This contrast will be still higher in the nuclear region, provided
that a significant fraction of the CO emission in Mrk 231 arises from a more
extended region. Thus the H$_2$O-to-CO line intensity ratios 
in the SPIRE wavelength range are an excellent diagnostic of
extragalactic compact/warm far-IR continuum sources with unusually high
amounts of H$_2$O.  

The above comparison also indicates that the nuclear region of Mrk 231 cannot
be interpreted as an ensemble of classical PDRs. Three main scenarios are
proposed to explain such high amounts of H$_2$O: $(i)$ widespread
shocks/cosmic rays: although the H$_2$O lines peak around the
systemic velocity, outflows
of $\sim100$ km s$^{-1}$ are not ruled out by our data, and indeed some
indications in the H$_2$O line shapes of systematic motions have been found;
an enhanced cosmic ray flux could also have an important impact on the nuclear
chemistry. 
$(ii)$ XDR chemistry: our derived H$_2$O abundance of
$\sim10^{-6}$ is in very good agreement with XDR model results by 
\citet[][their Fig. 3, Model 3]{mei05}, as well as with our preliminary
estimate of the H$_2$O spatial distribution.
$(iii)$ An undepleted chemistry, where H$_2$O that formed on grain mantles
is released into the gas phase, as in Galactic hot
cores; in support of this 
scenario, the derived $T_{\mathrm{dust}}$ in $W_C$ is close to the
evaporation temperature of solid H$_2$O. 
All three scenarios are probably taking place, and the identification
of the dominant process requires a multi-species analysis.

The nuclear region traced by the high-lying H$_2$O lines has a
size similar to the nuclear disk (or outflow) observed at radio wavelengths
and H I 21 cm by \citet[][their Figs. 3 \& 7]{car98}, 
suggesting a close physical correspondence. 
From $H_C$ and $W_C$, the nuclear surface brightness ($\sim1.5\times10^{13}$
\Lsun\ kpc$^{-2}$) exceeds the highest values attained in starburst on
spatial scales $\gtrsim100$ pc \citep{meu97,dav07}, while the nuclear
luminosity-to-mass ratio ($L/M\sim4\times10^3$ \Lsun/\Msun) exceeds the limit
for a starburst estimated by \citet{sco03}.
From near-IR data, \citet{dav07} estimated a starburst luminosity from a
similarly sized region of $\lesssim7\times10^{11}$ \Lsun; according to the 
joint luminosity of our $H_C$ and $W_C$, the AGN would account for at least
50\% of the output power in Mrk 231.

\begin{acknowledgements}
      We thank the SHINING consortium for proving us with the spectrum of the
      H$_2$O \t423312, E. Habart for providing us
      the SPIRE spectrum of the Orion Bar prior to its publication in this
      volume, and the SPIRE ICC FTS team for their great help in data
      reduction/analysis. E.G-A is a 
      Research Associate at the Harvard-Smithsonian Center 
      for Astrophysics. Dark Cosmology Centre is funded by DNRF.
\end{acknowledgements}

\end{document}